\documentclass{eas}
\usepackage{graphicx}
\usepackage[authoryear]{natbib}

\newcommand{\vc}[1]{{\boldmath #1}}
\begin{document}

\title{Modelling the evolution of planets in disks} 
\author{Wilhelm Kley}
\address{Institute of Astronomy \& Astrophysics,
University of T\"ubingen, Germany}
%
%
\begin{abstract}
To explain important properties of extrasolar planetary systems
(eg. close-in hot Jupiters, resonant planets) an evolutionary scenario
which allows for radial migration of planets in disks is required.
During their formation protoplanets undergo a phase in which they are
embedded in the disk and interact gravitationally with it.
This planet-disk interaction results in torques (through gravitational forces)
acting on the planet that will change its angular momentum and result
in a radial migration of the planet through the disk.
To determine the outcome of this very important process for planet
formation, dedicated high resolution numerical modeling is required.
This contribution focusses on some important aspects of the
numerical approach that we found essential for obtaining  successful results. 
We specifically mention the treatment of Coriolis forces, Cartesian grids,
and the {\tt FARGO} method.
\end{abstract}
\maketitle
\section{Introduction}
After the discovery of now over 300 extrasolar planets 
(for an always up to date list, see: {\tt http://exoplanet.eu/}
by Jean Schneider) the process of planet formation has now become
again a major area of modern astrophysical research.
The dynamical properties of the newly discovered planetary system
show distinct differences with our own Solar System,
see \citet{2007prpl.conf..685U} for a review. Most noticeable are
very close in massive planets (hot Jupiters and Neptunes), high orbital
eccentricities, and the occurrence of low order (2:1) mean-motion resonances.
Some of these properties require an orbital evolution of the embedded planets
through the disk. This migration process is very generic for young planets
that are still embedded in the disk 
\citep{1980ApJ...241..425G,1986ApJ...309..846L,1997Icar..126..261W}.

The formation of massive planets can take place primarily along two routes:
({\it i)} through {\it gravitational instability} of the protoplanetary disk
\citep{1998ApJ...503..923B}.
If the disk is sufficiently massive (about $0.1 M_*$) spiral density arms
will form which may turn gravitationally unstable resulting
in local fragmentation and the formation of high density 'blobs',
i.e. the protoplanets. The main advantage of this scenario is its speed,
as it is possible to form Jupiter sized planets within 100 orbital
periods. A possible drawback is the requirement of relatively rapid cooling
mechanisms in the disk, since the fragments can only collapse if they can cool
sufficiently rapidly. Also, the existence of massive solid cores in the centers
of the planets cannot easily be explained.
({\it ii)} through the {\it core accretion} process \citep{1980PThPh..64..544M}.
In this second scenario the
early growth of planets is accomplished by coagulation of small dust particles which
are embedded in the gaseous protoplanetary disk. They collide, stick together
and grow in mass until eventually (after several stages) a solid core of a few earth
masses has be assembled. At this stage, gas accretion sets in and the planet can 
grow in mass up to several Jupiter masses.

\begin{figure}[ht]
\centerline{
\resizebox{0.65\linewidth}{!}{%
\includegraphics{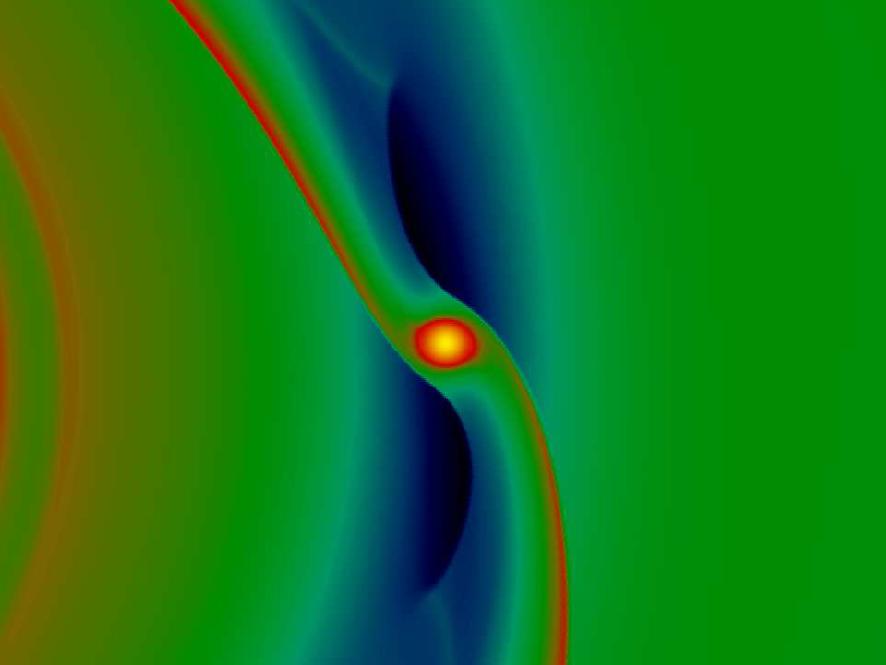}}
}
  \caption{
Density structure of an embedded Saturn mass planet in a protoplanetary disk. Clearly
visible are the two spiral arms (red) and the lower density (gap, dark blue)
at the orbit of the planet.
(Courtesy: Aur{\'e}lien Crida)
  }
   \label{kley:fig-crida}
\end{figure}

Both models of planet formation must take place within a gaseous environment,
i.e. the planet is still embedded within the surrounding protoplanetary
accretion disk.
In this case, the presence of the growing protoplanet in the disk
will generate non-axisymmetric disturbances, the spiral arms (see Fig.~\ref{kley:fig-crida}).
These pull gravitationally on the planet or, phrased differently, exert a
torque $\vc{T}$ on the planet. As a consequence the angular momentum $\vc{L}$
of the planet will be changed $\dot{\vc{L}} = \vc{T}$, and since for
circular orbits $\vc{L}$ is only a function
of the semi-major axis $a$ of the planet, the planet has to migrate.
Hence, in both planet formation scenarios,
the planet will move radially through the disk.
Two observational facts are typically taken as indirect evidence that such a planetary
migration process has indeed occurred. First, the existence of hot planets close
to the star, as in-situ formation is difficult, and second the
occurrence of mean motion resonances, as they require special locations of the planets
which are very unlikely to occur just by chance.
To calculate the evolution, i.e. the mass growth and migration of young planets in the
disk, numerical simulations are typically employed. In this contribution we focus
on the numerical aspect of planet-disk modeling, and will
outline some of the necessary requirements to perform those simulations successfully.
\section{Modelling}
The majority of disk models assume a vertically thin disk which is approximated
by an infinitesimally thin disk lying in the equatorial ($x-y$) plane. The motion of
the gas is mostly described making a viscous hydrodynamic approach. Hence, the equations
to be solved consist of the two-dimensional Navier-Stokes equations. Additionally the
gravitational potentials of the central star and the embedded planets have to be added.
The full equations including the viscous terms are given in \citet{1999MNRAS.303..696K}.
Initially, purely two-dimensional hydrodynamical simulations of this problem have
been performed by several groups 
\citep{1986ApJ...307..395L, 1999MNRAS.303..696K,
1999ApJ...514..344B, 1999ApJ...526.1001L,2000MNRAS.318...18N,
2002A&A...387..605M,2002A&A...385..647D,2006Icar..181..587C}, subsequently extended
to three-dimensions \citep{2001ApJ...547..457K}, and also with nested grids
\citep{2003ApJ...586..540D}. Later full
3D-MHD simulations with embedded planets have
been added \citep{2003MNRAS.339..993N,2003ApJ...589..543W}.
All of these initial models used a locally isothermal equation
of state simplifying the numerical modelling. Through fully
3D simulations including radiative effects,
\citet{2006A&A...459L..17P} demonstrated the
crucial role the thermodynamics can play in determining the migration,
an area that gains presently a lot of momentum
\citep{2008A&A...478..245P,2008ApJ...672.1054B,2008A&A...487L...9K}.
However, in this contribution we shall leave these most recent developments aside
and focus in particular on some important numerical aspects of the planet-disk
problem.

\begin{figure}[ht]
\begin{minipage}{0.60\linewidth}
\includegraphics[height=.30\textheight]{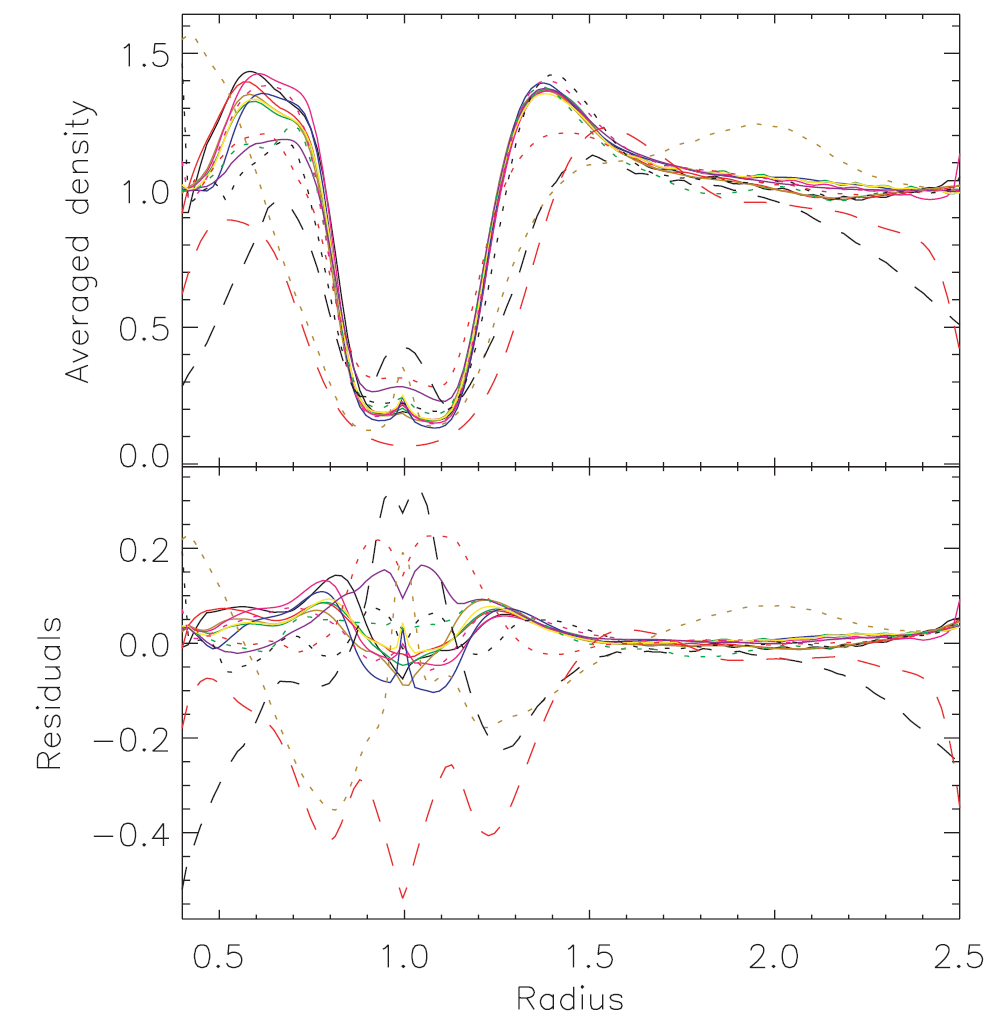}
\end{minipage}
\begin{minipage}{0.40\linewidth}
\includegraphics[height=.30\textheight]{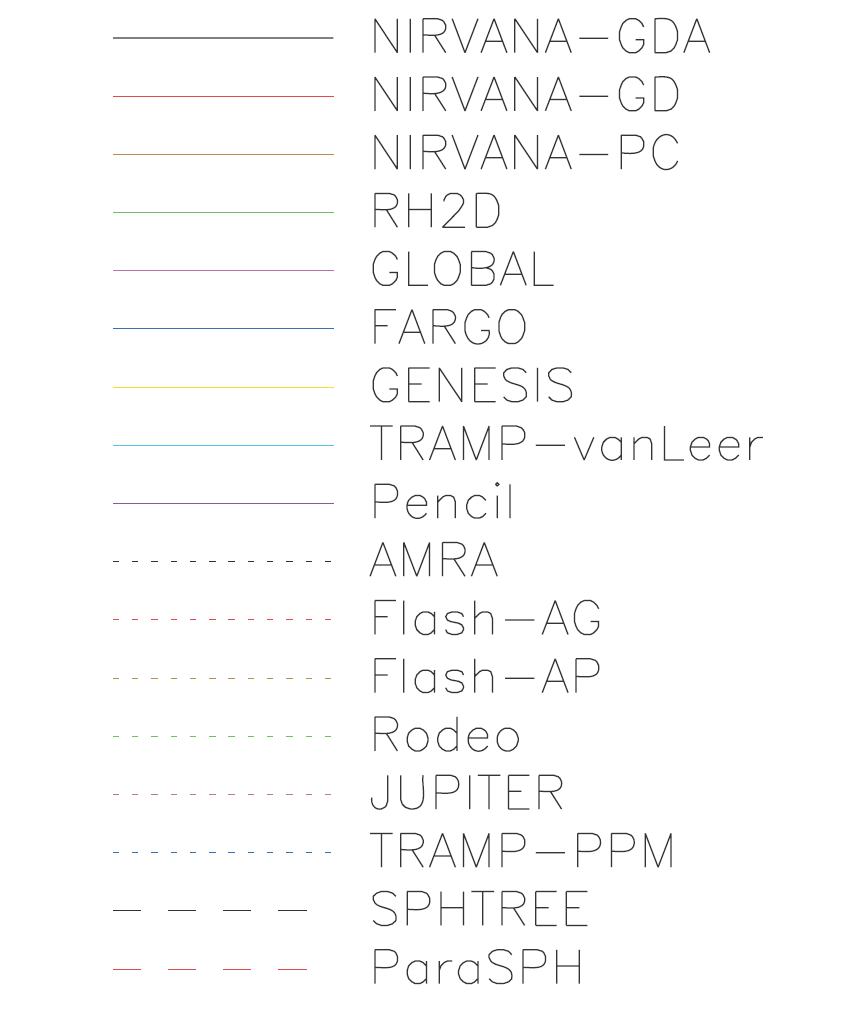}
\end{minipage}
\caption{
Azimuthally averaged density profile for an embedded Jupiter mass planet in a
viscous ($\nu=10^{-5}$) and locally isothermal ($H/r=0.05$) disk after 100 orbital periods.
Displayed are the results for different codes as given in the legend. Here the solid
lines refer to upwind-codes (with the exception of the 
last, purple curve which refers to the high-order code
{\tt PENCIL} \citep{2002CoPhC.147..471B}), dotted lines to Riemann solver,
and dashed lines to SPH codes. (Adapted from: \citet{2006MNRAS.370..529D})
  }
  \label{kley:fig-eucomp}
\end{figure}

A standard setup for planet-disk modelers has been described recently within a European-wide
code-comparison project where 22 Co-authors have used about 15 different codes
on the same physical problem \citep{2006MNRAS.370..529D}. 
In the comparison paper grid codes (upwind, Riemann), particle codes (SPH) and cylindrical or
Cartesian grids are compared against each other.
Anyone interested in
performing planet-disk simulations is strongly encouraged to test and compare his/her results
with that paper. One major simplification that is typically applied is the usage of an
(locally) isothermal equations of state without any energy equation.
For this case there exist also semi-analytical linear results for small mass planets
\citep{1980ApJ...241..425G, 1997Icar..126..261W}.
The principle outcome of such simulations is displayed in Fig.~\ref{kley:fig-crida},
which is based on a grid based simulation using a cylindrical coordinate system.
An important quantity to plot is the radial profile of the azimuthally averaged
surface density in the disk.
This gives an indication of the accuracy of the angular momentum conservation and transport
in the disk, which is particularly important to calculates reliable migration rate.
Results obtained in the comparison project are displayed Fig.~\ref{kley:fig-eucomp} for
the density. While most of the codes agree, there are also important deviations.
Let us concentrate on some important points in performing these simulations. 
\subsection{Coriolis terms}
Often it is desirable to perform the simulations in a coordinate frame that corotates with
the planet. In such a case additional source terms appear in the equations of motion.
Taken at face value these additions imply non-conservation for the
angular momentum. Through a reformulation of the $\varphi$ equation one
can write an explicit conservation equation for the
angular momentum (as viewed in the inertial frame). Only the usage of this
conserving scheme yields the correct density distribution and reliable
torques acting on the planet, as demonstrated by \citet{1998A&A...338L..37K}.
A non-conservative scheme will typically
require a much higher spatial resolution for obtaining equally good results.
\begin{figure}[ht]
\begin{minipage}{0.50\linewidth}
\includegraphics[height=.30\textheight]{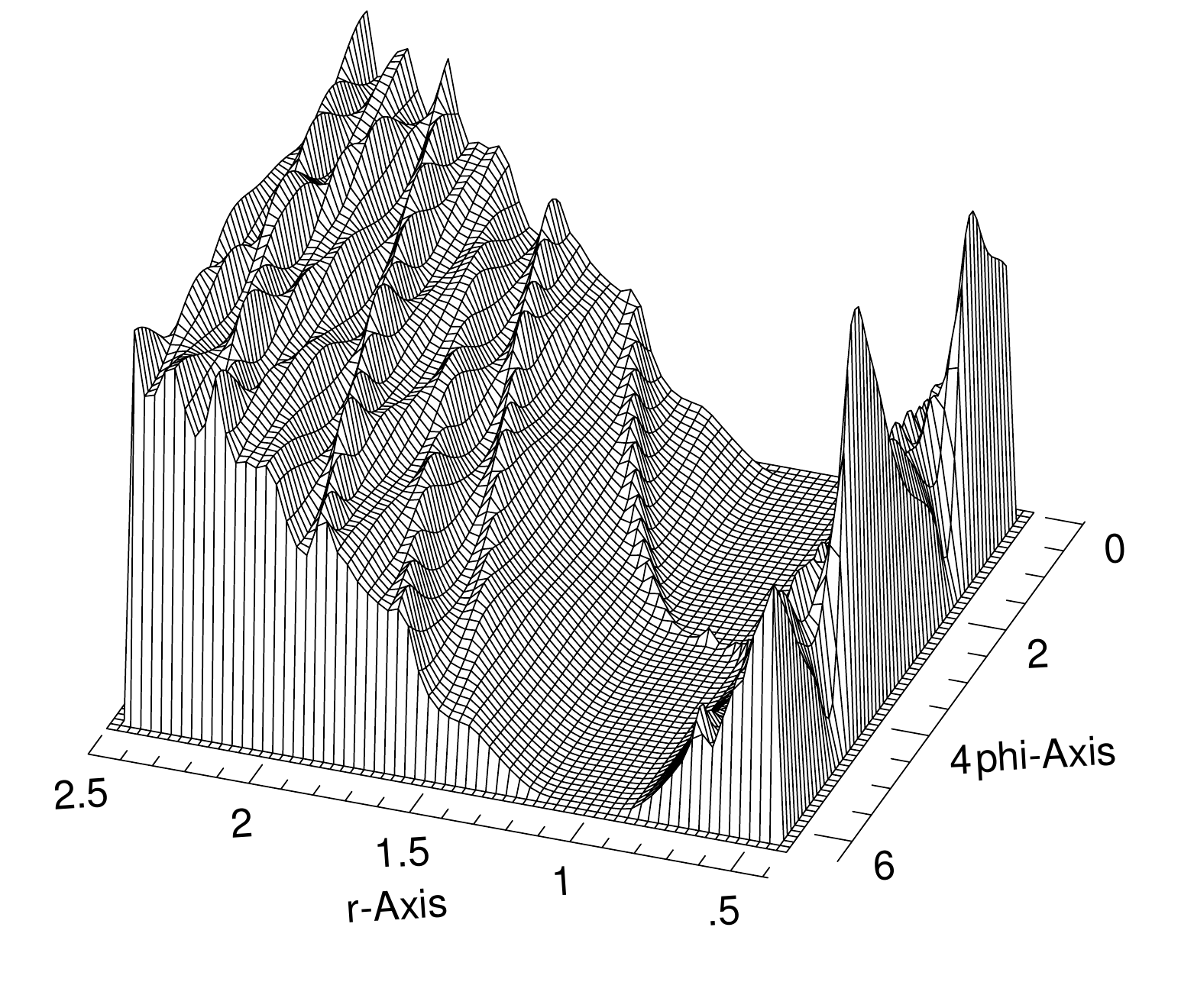}
\end{minipage}
\begin{minipage}{0.50\linewidth}
\includegraphics[height=.30\textheight]{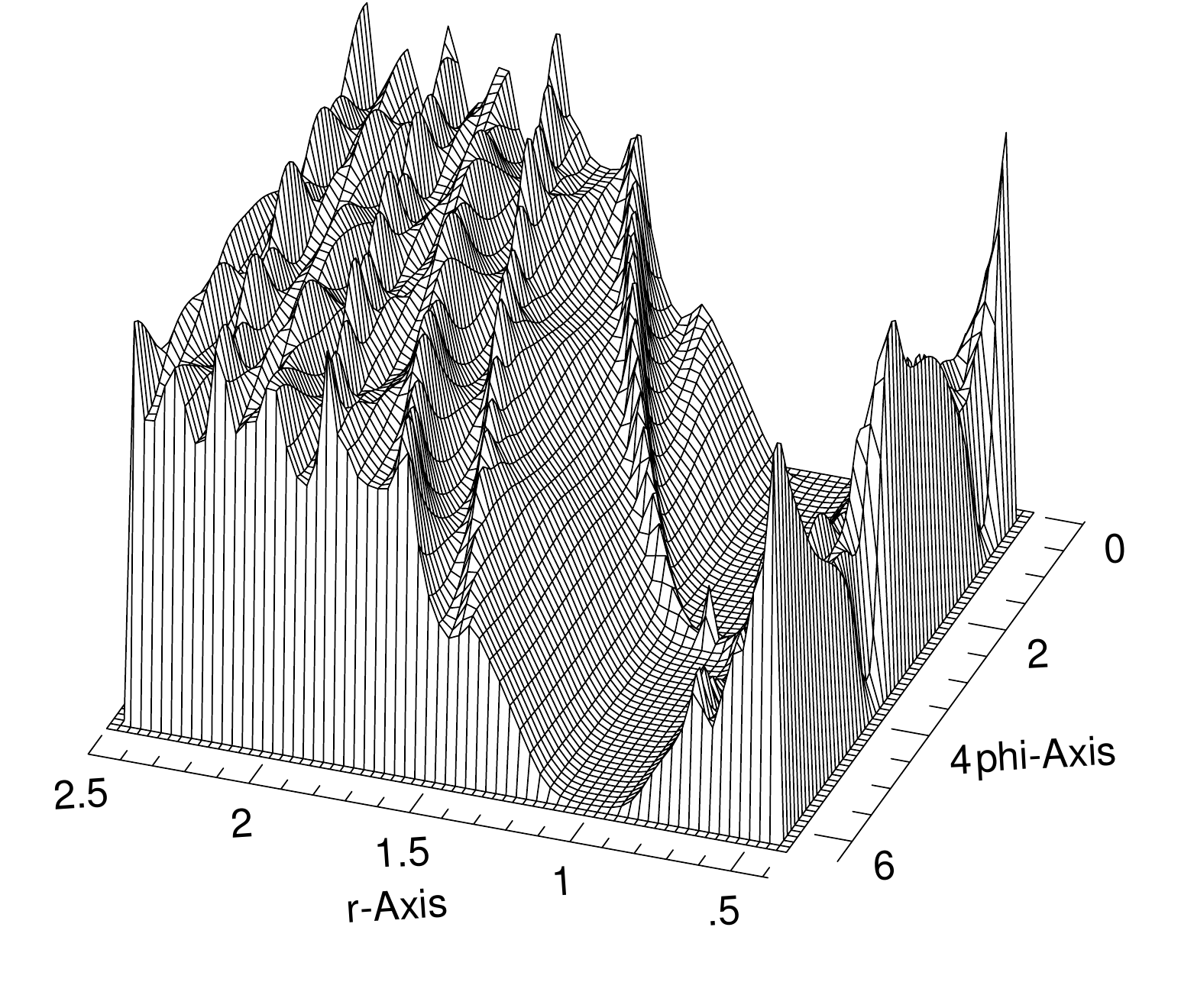}
\end{minipage}
\caption{
3D density distribution of the perturbed disk in the presence of
an embedded planet calculated in the rotating frame. 
Left: explicit sources, non-conservation of angular momentum.
Right: conservative formulation.
(Adapted from: \citet{1998A&A...338L..37K})
  }
  \label{kley:fig-coriol}
\end{figure}
The difference of the two formulations is demonstrated in
Fig.~\ref{kley:fig-coriol} where a Jupiter mass planet has been embedded in an
accretion disk. The density distribution, in particular the slope near to the
planet is very different in the two cases. The right panel refers to the conservative
formulation and the obtained results are very close to the inertial
case. The incorrect density gradient in the vicinity of the planet
for the non-conservation case leads to wrong migration and mass accretion rates.
Hence, when using a rotating reference frame, care should be taken to
always write the angular momentum in conservative form.
\subsection{The {\tt FARGO} algorithm}
\begin{figure}[ht]
\begin{minipage}{0.50\linewidth}
\includegraphics[height=.25\textheight]{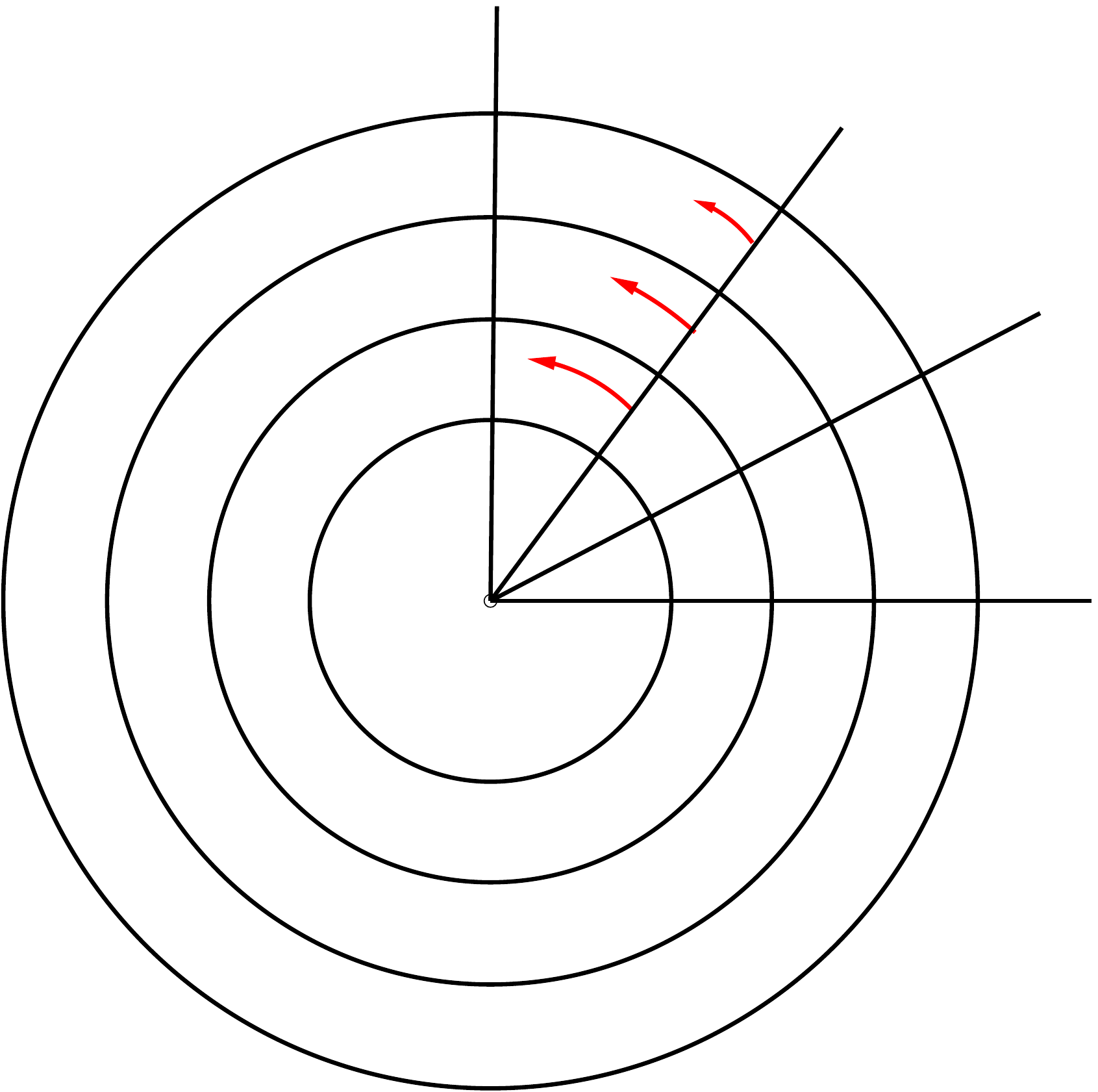}
\end{minipage}
\begin{minipage}{0.50\linewidth}
\includegraphics[height=.25\textheight]{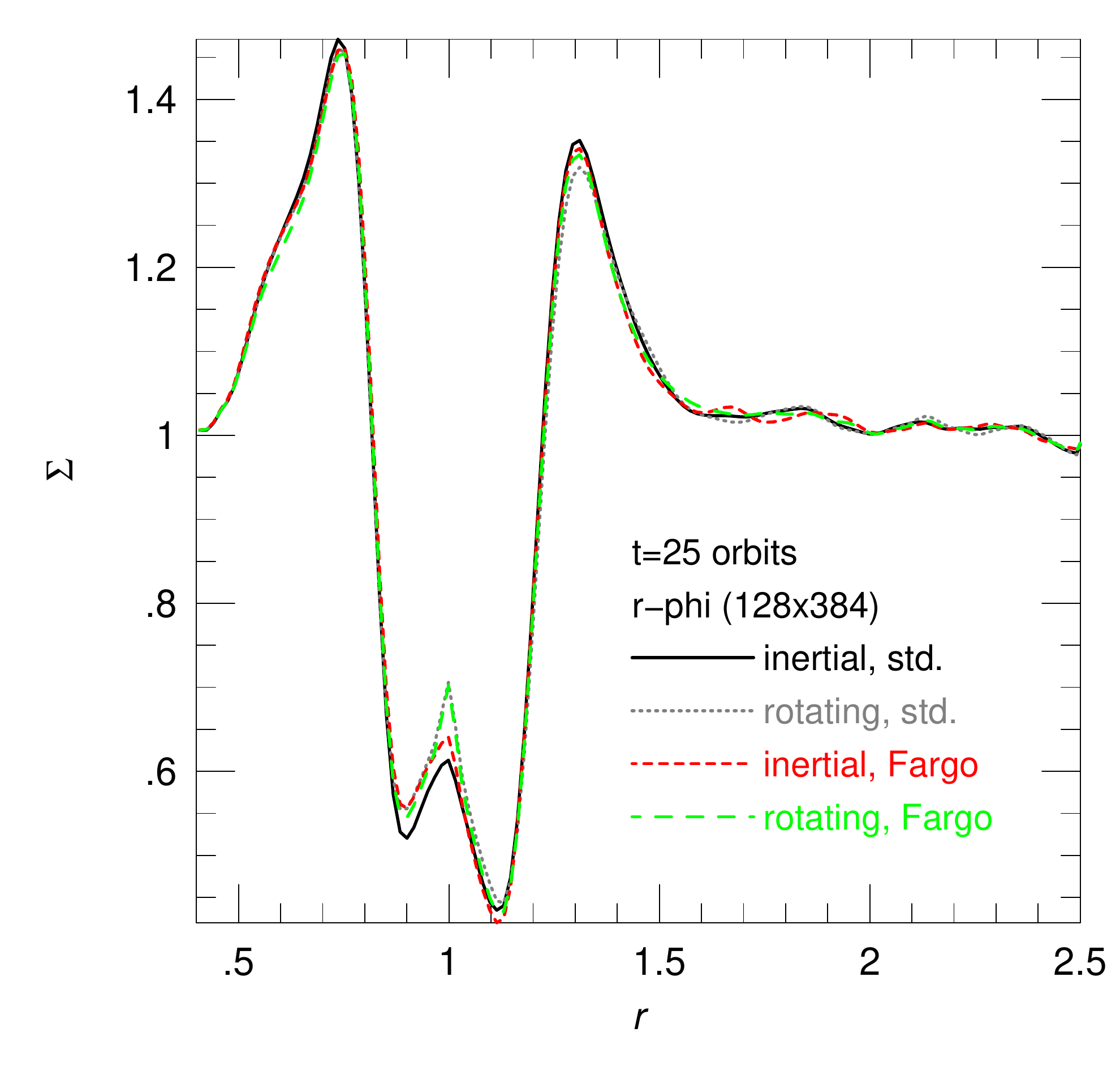}
\end{minipage}
\caption{
{\bf Left} Illustration of the angular velocity in a Keplerian disk.
The large inner velocities imply a very small timestep.
{\bf Right} The azimuthally averaged density for four kinds of
simulations using the standard method
and the {\tt FARGO}-implementation \citep{2000A&AS..141..165M}
both in an inertial and rotating reference frame.
  }
  \label{kley:fig-fargo}
\end{figure}

Typically, the simulations as just described are performed in a cylindrical coordinate
system ($r, \varphi$). In explicit codes the time step is limited by
the Courant condition which states that in one time step the information can only
be propagated over at most one gridcell to ensure numerical stability.
In a highly supersonic accretion disk
where the flow is basically on circles this yields
\begin{equation}
  \Delta t  <  \frac{\Delta \varphi}{\omega}
\label{kley:eq-dt}
\end{equation}
where $\omega$ denotes the azimuthal angular velocity and $\Delta \varphi$
the azimuthal size of a gridcell.
In a Keplerian disk  $\omega$ scales as $\omega \propto r^{-3/2}$ 
and hence the timestep scales as $\Delta t \propto r^{3/2}$, i.e.
the innermost rings with the smallest radii determine the size of the timestep,
see also Fig.~\ref{kley:fig-fargo} left panel.
In our case of planet disk-interaction the inner disk boundary must not
be very close to the planet and, due to the Courant condition, this leads to
very small timesteps. An excellent remedy of this problem has been suggested
by \citep{2000A&AS..141..165M} who introduced the so-called
{\tt FARGO}-algorithm (Fast Advection in Rotating Gaseous Objects). 
This method relies on a directional splitting of the advection,
where first the radial advection is performed in the standard way.
The azimuthal part is then done in two parts:
First an average
angular velocity $\bar{\omega}_i$ is calculated for each ring $i$, and
all quantities at each ring $i$ are shifted by an integer number of
gridcells, $n_i$={\tt Nint}$(\bar{\omega}_i \Delta t/\Delta \varphi$),
where {\tt Nint} denotes the nearest integer function.
This corresponds to a transport by the 'shift velocity' 
${\omega}_i^{SH} = n_i \Delta \varphi/\Delta t$.
In the second step, all quantities are transported (in two sub-steps)
with the remaining part of the angular velocity 
${\omega}_i^{r}= {\omega}_i - {\omega}_i^{SH}$
using the standard advection routine at hand.
In this case the effective applied transport velocity
(${\omega}_i^{r}$) in each ring is of the order of the deviations from the
mean $\bar{\omega}_i$ which is much smaller than the Keplerian flow speed.
Using this method, a large increase in the time-step can be achieved.
Since the first shift-stip is exact, there is also a substantial reduction
in diffusivity of the method. For details of the implementation see
\citet{2000A&AS..141..165M,2000ASPC..219...75M}. The code with all sources and improvements
is available at the 
{\tt FARGO}-webpage: {\tt http://fargo.in2p3.fr/}.
In Fig.~\ref{kley:fig-fargo} we demonstrate in the right panel the accuracy of
the method by comparing 4 different integration methods. Calculated are runs for
the inertial and rotating frame with either {\tt FARGO} switched or not. Here, the
disk extends radially from $[0.4;2.5]$ and the grid is covered by
$128\times384$ gridcells. This setup refers to the standard test case as
described above \citep{2006MNRAS.370..529D}. Displayed are the
results after 25 orbits of the planet.
Clearly all four cases yield very similar results. The runs were all done with
a Courant number of $0.75$, and the used number of timesteps for the runs from top down
have been: 51,000; 39,000; 5800; 5800. The first case (standard, inertial)
needs the most timesteps, the second (standard, rotating) needs about 20\% less
due the the reduced angular velocities in the rotating frame. The last
two {\tt FARGO}-runs yield both (in the inertial and rotating frame) an
identical number of timesteps, since the introduction of the shift brings
all the rings essentially to the corotating frame.
As seen in this case the speedup obtained with  {\tt FARGO} is about a factor
of 9! In the general case the speedup depends on the radial range
and the grid scaling. For a logarithmic radial grid an even larger
speedup factor can be achieved. Obviously the {\tt FARGO}-method
gives the correct results and hence should definitely be the favored
method when performing simulations of sheared flow in disks.
\subsection{Coordinate System}
\label{kley:subsec-coord}
As mentioned above, typical planet-disk simulations are performed using
a cylindrical coordinate system, which is well adapted to the physics
of the problem and yields automatically the important feature of
angular momentum conservation.
However, recently several simulations have been presented that utilize
a Cartesian coordinate system on the planet-disk problem 
\citep{2008ApJ...676..639Z, 2008MNRAS.386..164P, 2008arXiv0810.3192L}.
This has inspired us to perform some test simulations in Cartesian
coordinates as well, and evaluate the accuracy and requirements.
For this purpose we used our standard code {\tt RH2D}, which is a grid
based code utilizing a 2nd order upwind scheme (monotonic transport), staggered
grid and operator splitting \citet{1989A&A...208...98K}.
The code behaves quite similar to the well known ZEUS
code. What we found first is that in the standard formulation, where advection
is operator-split from the force terms, disk calculations 
(even without embedded planet) in Cartesian coordinates
were not possible. We believe that this failure is due to the usage of
operator splitting with the separation of advection and forces.
Here, the disk structure is given by the equilibrium of 
gravity and inertial terms which are both large, have to cancel out,
but are not done in one step.

A reformulation of the numerical scheme using a 2nd order TVD Runge-Kutta
time-integrator \citep{1988JCoPh..77..439S}
and no operator-splitting but otherwise identical method
(staggered, monotonic transport) gave consisted results on disk evolution calculations.
\begin{figure}[ht]
\centerline{
\includegraphics[height=.30\textheight]{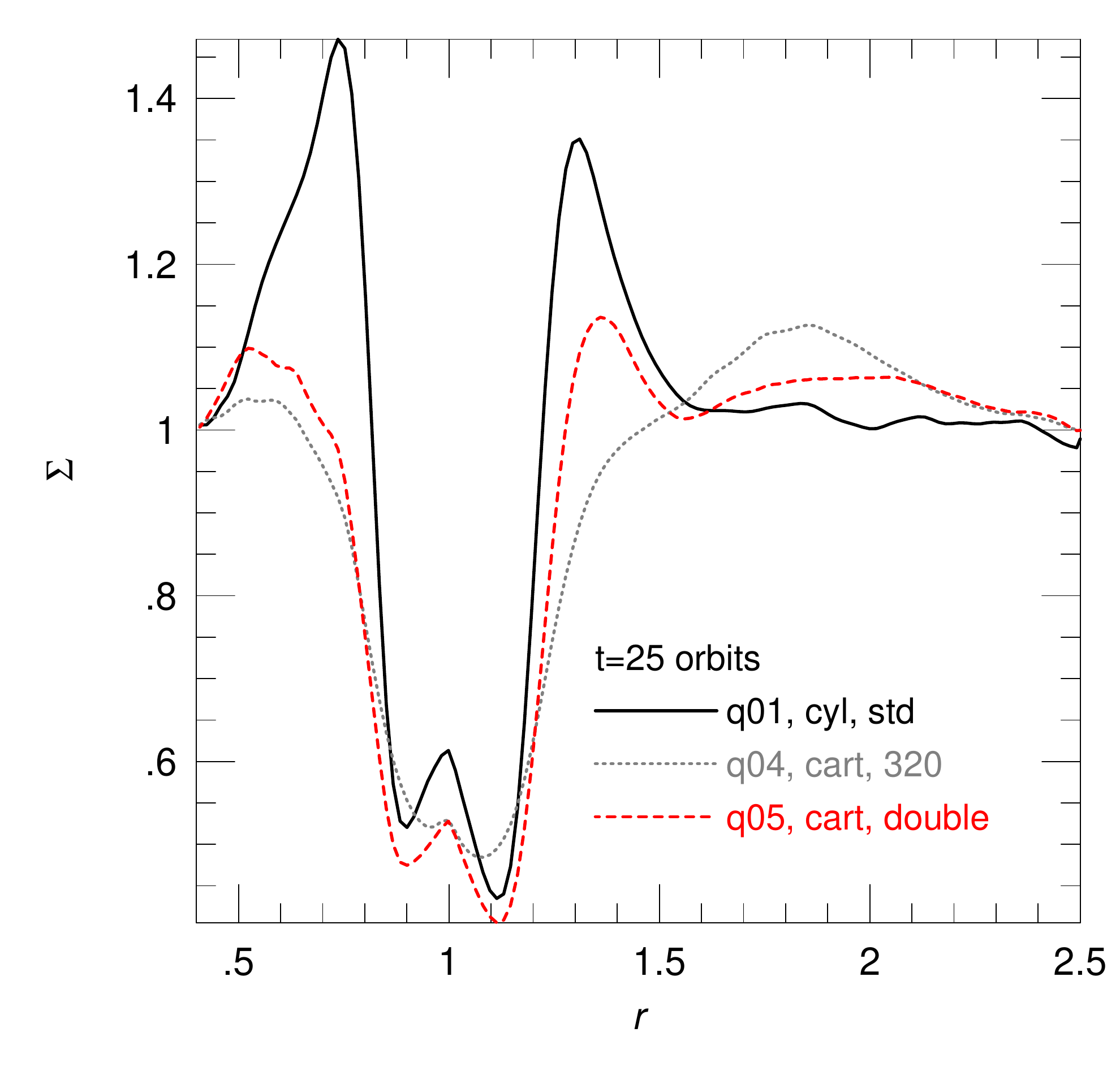}
}
\caption{
The azimuthally averaged density for different simulations
using a Cartesian grid structure at two different resolutions
(grey dotted line: $320\times 320$, red dashed line:
$640\times 640$) or a cylindrical grid (black solid: $128\times 384$).
The simulations refers to the same physical setup as above in Fig.~\ref{kley:fig-eucomp}.
  }
  \label{kley:fig-cartesian}
\end{figure}
The results for the standard problem after 25 orbits are displayed in
Fig.~\ref{kley:fig-cartesian} for two Cartesian runs with $320\times320$
and $640\times640$ gridcells in a square with length $[-2.5,2.5]$ in the
$x$ and $y$ direction. As boundary conditions for these Cartesian runs
we used the same damping conditions as in \citet{2006MNRAS.370..529D}.
The smaller grid $320\times320$ yields in the vicinity of the planet
the same resolution as an $128\times 384$ grid in cylindrical coordinates,
which is also displayed in Fig.~\ref{kley:fig-cartesian}.
While the depth of the gap created by the planet is similar in all
cases, its width varies considerably. The Cartesian runs do not seem to have
converged even in the high resolution case.
In the EU-comparison project, there have been two codes that used
a Cartesian grid: The {\tt PENCIL} code (purple line in Fig.~\ref{kley:fig-eucomp})
and the {\tt FLASH} code (brown dashed line, labeled `Flash-AP'). Of these
two, only the {\tt PENCIL} code gave satisfactory results, where the gap
width shows good agreement, but the gap is a little to shallow.
On the other hand, the gap width is much larger for the {\tt FLASH}
code using a Cartesian coordinate system. Only the two SPH-codes (long dashed
lines) show worse results in this test case. For the SPH-method, there
seems to be a possible remedy which is currently tested (R.~Speith,
private communication).
From these Cartesian simulations we conclude that {\it i)} due to the non-conservation
and enhanced diffusion of angular momentum, a much higher resolution than in 
cylindrical codes is required for similar results, or {\it ii)} very
high order schemes such as {\tt PENCIL} have to be applied.
\begin{figure}[ht]
\centerline{
\includegraphics[height=.25\textheight]{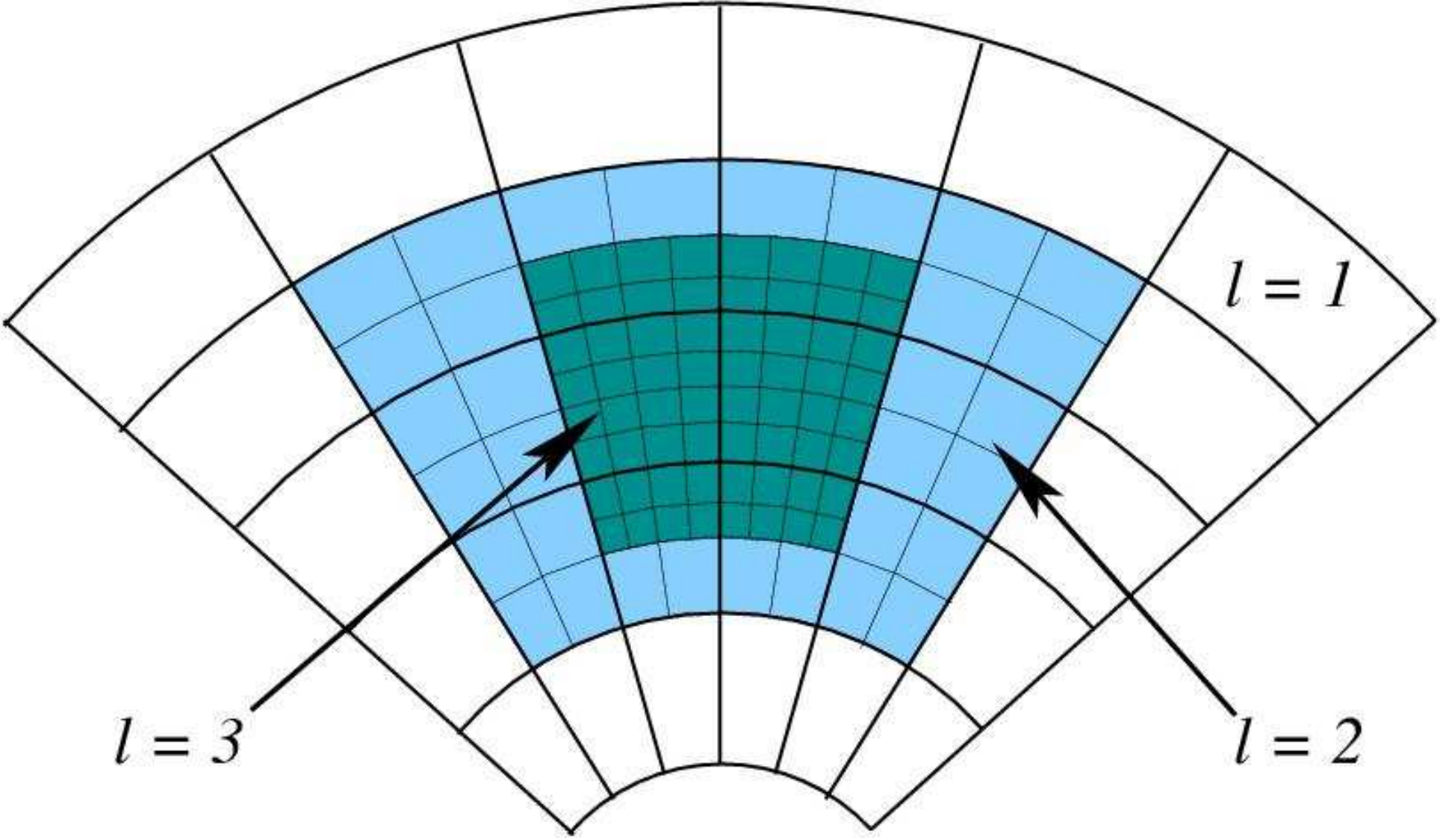}
}
\caption{
The grid structure of a nested grid with 2 levels of refinements.
Around the fixed location of the planet (here assumed to be in the
very center of the grids) the base grid ($l=1$) is covered by superimposed
finer grids ($l=2,3$) to locally enhance the resolution. These grids are already in place
from the very beginning of the simulations and not added later.
(Adapted from: \citet{2002A&A...385..647D})
  }
  \label{kley:fig-nested}
\end{figure}
\subsection{Nested Grids}
Another important numerical issue refers to the fact that for an accurate calculation
of the torques acting on the planet, the local as well as global structure of
the disk is required simultaneously. To obtain this with one single grid would require
a very large resolution indeed. A way out is to refine the grid structure in the vicinity
of the planet. This has been attempted through the usage of a non-uniform spacing of the grid
\citep{2003MNRAS.341..213B}, which comes with the disadvantage of a large deformation of 
the gridcells near the planet. Another option is the usage of {\it Nested Grids} which
has been utilized by \citet{2002A&A...385..647D}.
In such a case the base grid is refined locally at locations that are predefined 
(fixed) from  the very beginning of the simulations, in contrast to adaptive mesh
refinement (AMR) simulations. In the planet-disk interaction, frequently a non-moving
(fixed) planet is assumed to simplify the equations and focus directly on the
torque calculations. Here, the {\it Nested Grid} method is particularly well suited, as
the planet can be placed naturally in the center of the whole grid-system,
see Fig.~\ref{kley:fig-nested}, for the whole time evolution.
The {\it Nested Grid} technique has been used first in 2D \citep{2002A&A...385..647D} and
later fully 3D simulations of embedded planets in disks \citep{2003ApJ...586..540D}, where
up to 7 sub-grids have been used which has allowed for the most detailed resolution of the
Roche lobe of the protoplanet so far. Migration and mass accretion rates have been determined
through these simulations and it has been found that non-linear effects begin
to set in already at very small planetary masses, which has been confirmed later
by \citet{2006ApJ...652..730M}. To follow the planetary migration in a globally evolving disk
with a large radial range,
an interesting method has been suggested recently by \citet{2007MNRAS.377.1324C}
who sandwich the active 2D region with one-dimensional (only radial grids) added at the
inner and outer radius.
%
\subsection{Summary}
In this contribution we have discussed several important numerical issues related to
planet-disk simulations. We have shown that, due to the physical symmetry of the problem,
the usage of a cylindrical coordinate system is generally advantageous over a Cartesian
one. In case of a rotating coordinate system the equations need to be reformulated to
ensure angular momentum conservation. We have shown that a computational reformulation
of the angular advection using the {\tt FARGO} method leads to a substantial increase in the 
allowed size of the time-step. 
We have demonstrated the accuracy and achievable speed-up
of the {\tt FARGO} method by comparing the results
to standard cases. Finally, we have shown that an increase in resolution near to
the planet can been achieved through the usage of nested grids.
Future numerical models of the planet-disk problem will involve three-dimensional
analyzes which will include more physical effects such as self-gravity, radiation
transport and magnetic fields.
\subsection{Acknowledgements}
This work has been sponsered in parts by the DFG grants
KL-650/6, KL-650/7 and the Forschergruppe {\it The Formation of Planets.
The critical first growth phase} through grant KL-650/11.
\bibliographystyle{astron}
\bibliography{kley8}
\end{document}